\documentclass[aps,print,twocolumn,hidepacs,hidekeys]{revtex4}%
\usepackage{amssymb}
\usepackage{amsfonts}
\usepackage{amsmath}
\usepackage{graphicx}%
\providecommand{\U}[1]{\protect\rule{.1in}{.1in}}

\begin{document}
\title{Airy plasmons in graphene based waveguides}
\author{Rujiang Li$^{1,2,3}$, Muhammad Imran$^{1,2,3}$, Xiao Lin$^{1,2,3}$,
and Hongsheng Chen$^{1,2,3}$}
\affiliation{$^{1}$State Key Laboratory of Modern Optical Instrumentation, Zhejiang
University, Hangzhou 310027, China }
\affiliation{$^{2}$College of Information Science and Electronic Engineering, Zhejiang
University, Hangzhou 310027, China }
\affiliation{$^{3}$The Electromagnetics Academy of Zhejiang University, Zhejiang
University, Hangzhou 310027, China}
\email{(H. Chen): hansomchen@zju.edu.cn}

\begin{abstract}
In this paper, we propose that both the quasi-transverse-magnetic (TM) and
quasi-transverse-electric (TE) Airy plasmons can be supported in
graphene-based waveguides. The solution of Airy plasmons is calculated
analytically and the existence of Airy plasmons is studied under the paraxial
approximation. Due to the tunability of the chemical potential of graphene,
the self-accelerating behavior of quasi-TM Airy plasmons can be steered effectively,
especially in multilayer graphene based waveguides.
Besides the metals, graphene provides an additional platform to investigate
the propagation of Airy plasmons and to design various plasmonic devices.

\end{abstract}
\maketitle



\section{Introduction}

Quantum-optical analogies have always been the hot research topics in recent
years \cite{analogy,LPR3-243}. The concept of ``Airy beams'' is first proposed
within the context of quantum mechanics, where the probability density of the
Airy wave packet propagates without spreading, and accelerates even though no
force acts \cite{AJP}. Optics provides a fertile ground to realize the Airy
beams. To extend the concept of ``Airy beams'' into optics, the amplitude must
be truncated to ensure the containment of finite energy
\cite{OL32-979,OL32-2447} and to enable the experimental realizations
\cite{PRL99-213901}. In optics, Airy beams are non-diffracting waves that
propagate along a parabolic trajectory in a homogeneous medium. Besides, Airy
beams have the remarkable property of self-healing, where the beams tend to
reform during propagation after the action of perturbations \cite{OE16-12880}.
Due to their intriguing properties, the concept of Airy beams have been
extended into various fields of physics, such as temporal pulses
\cite{nphoton4-103,OE19-2286}, spin waves \cite{PRL104-197203}, plasma
\cite{science324-229}, and water waves \cite{PRL115-034501}.

Airy plasmons are surface plasmon polaritons (SPPs) that propagate along the
metal-dielectric interface in the form of Airy beams \cite{LPR8-221}.
Considering the subwavelength confinement of SPPs, Airy plasmons are very
attractive for ultra-compact planar plasmonic devices. To insure the validity
of the paraxial approximation and a large enough self-deflection distance,
Airy plasmons can only exist in certain parameter ranges \cite{OL35-2082}. For
the metal-based plasmonic waveguides, the generation
\cite{PRL107-116802,OL36-3191, PRL107-126804}, collision \cite{OL37-3402}, and
control \cite{OL36-1164,OL38-1443} of Airy plasmons have all been studied.

In recent years, graphene, a two dimensional hexagonal crystal carbon sheet
with only one atom thick, has attracted much attention as a good counterpart
in the THz frequencies of metals in the optical frequencies \cite{NL11-3370}.
Since its ballistic transport and ultrahigh electron mobility \cite{RMP81-109,
RMP82-2673,RMP83-407}, the surface conductivity of graphene is almost purely
imaginary in the THz frequencies \cite{science332-1291}. In other words,
graphene can be treated as a thin film of metal with a low loss. Due to this
intriguing property, Airy plasmons have been proposed in a plasmonic waveguide
based on the monolayer graphene without the paraxial approximation
\cite{ieeepj}. However, strictly speaking, Airy beam is a solution of the
Schr\"{o}dinger equation in the paraxial approximation. The validity of the
paraxial approximation and the properties of Airy plasmons in graphene based
waveguides should be discussed. In this paper, based on the analytical
solution of Airy plasmons in the paraxial approximation, we will show that
both the quasi-transverse-magnetic (TM) and quasi-transverse-electric (TE)
Airy plasmons can be supported in graphene-based waveguides. Compared with the
metal based waveguides, the self-accelerating behavior of Airy plasmons in
graphene based waveguides can be steered effectively by tuning the chemical
potential of graphene.

This paper is organized as follows. First, in section II, we briefly
review the plasmonic modes of graphene based waveguides. Second, the
Airy plasmons supported in graphene based waveguides are discussed
in section III. Third, in section
IV, the self-accelerating behavior of Airy plasmons is steered effectively
both in monolayer and multilayer graphene based waveguides
by tuning the chemical potential of
graphene. Finally, section V is the conclusion.

\section{Plasmonic modes}

\begin{figure}[ptb]
\centering
\vspace{0.5cm} \includegraphics[width=6cm]{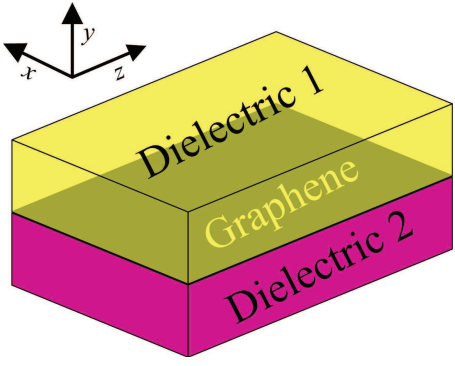} \vspace{-0.0cm}%
\caption{(Color Online) Schematic of the structure, where the graphene layer
is placed on the $xz$ plane, and sandwiched by the dielectric 1 (yellow area)
and dielectric 2 (pink area). The permittivities and permeabilities of the two
dielectrics are ($\varepsilon_{1}$, $\mu_{1}$) and ($\varepsilon_{2}$,
$\mu_{2}$), respectively. }%
\label{structure}%
\end{figure}

For completeness, in this section, we will briefly review the plasmonic modes
supported by the planar dielectric-graphene-dielectric waveguide structure. The
structure is schematically shown in Fig. \ref{structure}, where the graphene
layer is placed on the $xz$ plane, and sandwiched by the dielectric 1 (yellow
area) and dielectric 2 (pink area). The permittivities and permeabilities of
the two dielectrics are ($\varepsilon_{1}$, $\mu_{1}$) and ($\varepsilon_{2}$,
$\mu_{2}$), respectively. 

This planar waveguide supports both the TM and TE plasmonic modes
\cite{PRL99-016803,nanotechnology24-345203}, where the dispersion relation is%
\begin{equation}
\frac{\varepsilon_{1}}{k_{1}}+\frac{\varepsilon_{2}}{k_{2}}+i\frac{\sigma_{g}%
}{\omega\varepsilon_{0}}=0 \label{TM dispersion relation 1}%
\end{equation}
for the TM mode,%
\begin{equation}
\frac{k_{1}}{\mu_{1}}+\frac{k_{2}}{\mu_{2}}-i\sigma_{g}\omega\mu_{0}=0
\label{TE dispersion relation 1}%
\end{equation}
for the TE mode, $k_{1}=\sqrt{\beta^{2}-k_{0}^{2}\varepsilon_{1}\mu_{1}}$,
$k_{2}=\sqrt{\beta^{2}-k_{0}^{2}\varepsilon_{2}\mu_{2}}$, $\omega=2\pi f$ is
the angular frequency, $k_{0}=\omega/c$ is the wavenumber in free space,
$\sigma_{g}$ is the surface conductivity of graphene, and $\beta$\ is the
propagation constant. If both the dielectric 1 and dielectric 2 are air,
namely $\varepsilon_{1}=\varepsilon_{2}=1$ and $\mu_{1}=\mu_{2}=1$, the
dispersion relation for the TM plasmonic mode reduces to%
\begin{equation}
\beta=k_{0}\sqrt{1-\left(  \frac{2}{\eta_{0}\sigma_{g}}\right)  ^{2}},
\label{TM dispersion relation 2}%
\end{equation}
and the dispersion relation for the TE plasmonic mode reduces to%
\begin{equation}
\beta=k_{0}\sqrt{1-\left(  \frac{\eta_{0}\sigma_{g}}{2}\right)  ^{2}},
\label{TE dispersion relation 2}%
\end{equation}
where $\eta_{0}=\sqrt{\mu_{0}/\varepsilon_{0}}$ is the characteristic
impedance of free space.

Besides, the surface conductivity of monolayer graphene can be calculated
according to the Kubo formula $\sigma_{g}\left(  \omega,\mu_{c},\tau
,T\right)  =\sigma_{\text{intra}}+\sigma_{\text{inter}}$, where
\begin{equation}
\sigma_{\text{intra}}=\frac{ie^{2}k_{B}T}{\pi\hbar^{2}\left(  \omega
+i\tau^{-1} \right)  }\left[  \frac{\mu_{c}}{k_{B}T}+2\ln\left(  e^{-\mu
_{c}/k_{B}T}+1\right)  \right]  \label{intra}%
\end{equation}
is due to intraband contribution, and
\begin{equation}
\sigma_{\text{inter}}=\frac{ie^{2}\left(  \omega+i\tau^{-1} \right)  }%
{\pi\hbar^{2}}\int_{0}^{\infty}\frac{f_{d}\left(  -\varepsilon\right)
-f_{d}\left(  \varepsilon\right)  }{\left(  \omega+i\tau^{-1} \right)
^{2}-4\left(  \varepsilon/\hbar\right)  ^{2}}d\varepsilon\label{inter}%
\end{equation}
is due to interband contribution \cite{JAP103-064302,JP19-026222}. In the
above formula, $-e$ is the charge of an electron, $\hbar=h/2\pi$ is the
reduced Plank's constant, $T$ is the temperature, $\mu_{c}$ is the chemical
potential, $\tau=\mu\mu_{c}/e v_{F}^{2} $ is the carrier relaxation time,
$\mu$ is the carrier mobility which ranges from 1 000 $\text{cm}^{2}%
/(\text{V}\cdot\text{s})$ to 230 000 $\text{cm}^{2}/(\text{V}\cdot\text{s})$
\cite{ACSnano}, $v_{F}=c/300$ is the Fermi velocity, $f_{d}\left(
\varepsilon\right)  =1/\left[  e^{\left(  \varepsilon-\mu_{c}\right)  /k_{B}%
T}+1\right]  $ is the Fermi-Dirac distribution, and $k_{B}$ is the Boltzmann's
constant. In the following, we choose a moderate mobility of $\mu=$ 50 000
$\text{cm}^{2}/(\text{V}\cdot\text{s})$, $T=300$ K, and the chemical potential
$\mu_{c}$ is tuned between $0$ eV to $1$ eV \cite{JAP103-064302}. Due to the
finite carrier relaxation time, the surface conductivity of graphene can be
expressed as $\sigma_{g}=\sigma_{g,r}+i\sigma_{g,i}$, where the real part
$\sigma_{g,r}$ is related to the optical loss of graphene, and the imaginary
part $\sigma_{g,i}$ is related to the permittivity \cite{science332-1291}.
Correspondingly, the propagation constant can also be expressed as
$\beta=\beta_{r}+i\beta_{i}$ according to Eqs. (\ref{TM dispersion relation 2}%
)-(\ref{TE dispersion relation 2}), where the imaginary part $\beta_{i}$ is
related to the propagation length of the plasmonic mode with $L=1/2\beta_{i}$
\cite{maier}.

\begin{figure}[ptb]
\centering
\vspace{0.0cm} \includegraphics[width=8cm]{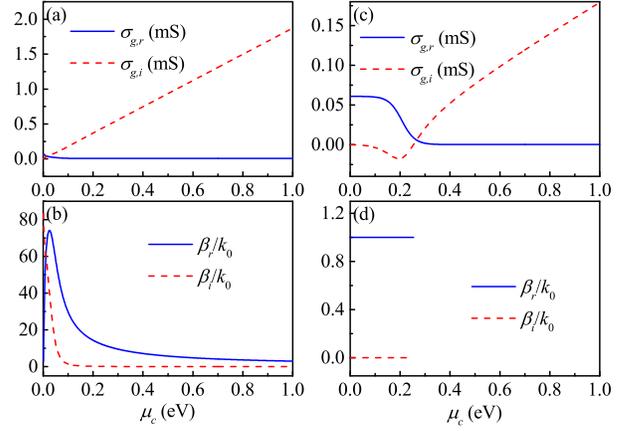} \vspace{-0.0cm}%
\caption{(Color Online) Top row: The dependence between the surface
conductivity and the chemical potential, where $\sigma_{g,r}$ is the real part
of surface conductivity, and $\sigma_{g,i}$ is the imaginary part. Bottom row:
The dependence between the propagation constant and the chemical potential,
where $\beta_{r}$ is the real part of propagation constant, and $\beta_{i}$ is
the imaginary part. The incident frequency is $f=10$ THz for (a)-(b) and
$f=100$ THz for (c)-(d).}%
\label{modes}%
\end{figure}

Based on Eqs. (\ref{intra})-(\ref{inter}), we can get the surface conductivity
of monolayer graphene at different frequencies. As shown in Fig.
\ref{modes}(a), for $f=10$ THz, both the real and imaginary parts of surface
conductivity are always positive. Since a positive $\sigma_{g,i}$ can lead to
a negative permittivity, this implies that this structure supports the TM
plasmonic mode, where the dependence between the propagation constant and the
chemical potential is shown in Fig. \ref{modes} (b). From the imaginary part
of surface conductivity, the propagation length of the TM plasmonic mode
decreases for small chemical potentials. In contrast, the dependence between
the surface conductivity and the chemical potential shows a different behavior
for $f=100$ THz, as shown in Fig. \ref{modes}(c). The imaginary part of
surface conductivity is negative for $0$ eV $\leq\mu_{c} \leq0.254$ eV, which
implies the existence of the TE plasmonic mode. Fig. \ref{modes}(d) shows the
dependence between the propagation constant and the chemical potential of the
TE mode, where the real part $\beta_{r}$ is nearly equal to the wavenumber
in free space, and the imaginary part $\beta_{i}$ is nearly equal to zero. Thus
the TE plasmonic mode supported by the planar waveguide is a weekly confined
mode that can propagate for a long distance.

\section{Airy plasmons}

In this section, we study the Airy plasmons that propagate in the planar
plasmonic waveguide shown in Fig. \ref{structure}. We assume that the solution
of Airy plasmons is a perturbation of the plasmonic mode, and the Airy
plasmons behave as a quasi-TM or quasi-TE mode approximately. The Helmholtz
equation is%
\begin{equation}
\nabla^{2}\Phi+k_{0}^{2}\varepsilon\left(  y\right)  \Phi=0, \label{Helmholtz}%
\end{equation}
where $\varepsilon\left(  y\right)  $ is the distribution of the relative
permittivity, $\Phi=H_{y}(x,y,z)$ for the quasi-TM Airy plasmons, and
$\Phi=E_{y}(x,y,z)$ for the quasi-TE Airy plasmons. The scalar field
$\Phi\left(  x,y,z\right)  $ can be expressed as a functional dependence of
the form
\begin{equation}
\Phi\left(  x,y,z\right)  =\psi\left(  x,z\right)  \Phi_{0}\left(  y\right)
\exp\left(  i\beta z\right)  , \label{electric field}%
\end{equation}
where the dimensionless scalar function $\psi\left(  x,z\right)  $ is
dependent on both the transverse direction $x$ and the propagation direction
$z$, the scalar function $\Phi_{0}\left(  y\right)  $ is only dependent on the
transverse direction $y$, and $\beta$ is a parameter that is related to the
$z$ component of the wavevector. Substituting Eq. (\ref{electric field}) into
Eq. (\ref{Helmholtz}), multiplying the result by $\Phi_{0}^{\ast}\left(
y\right)  $, and integrating over $y$ direction yields a scalar wave equation
\begin{align}
&  \left[  2i\beta\frac{\partial\psi}{\partial z}+\frac{\partial^{2}\psi
}{\partial x^{2}}\right]  +\left[  \frac{I_{2}}{I_{0}}+k_{0}^{2}%
\varepsilon-\beta^{2}\right]  \psi=0, \label{wave equation}%
\end{align}
where $I_{0}=\int_{-\infty}^{+\infty}\left\vert \Phi_{0}\left(  y\right)
\right\vert ^{2}dy$, $I_{2}=\int_{-\infty}^{+\infty}\Phi_{0}^{\prime\prime
}\left(  y\right)  \Phi_{0}^{\ast}\left(  y\right)  dy$, and the term
$\partial^{2}\psi/\partial z^{2}$ is neglected by employing the paraxial
approximation \cite{OL35-2082,OL32-674}. For the slowly varying amplitude,
the scalar function is
expressed as $\psi\left(  x,z\right)  =\phi\left(  x,z\right)  \exp\left\{
i\left[  I_{2}/I_{0}+k_{0}^{2}\varepsilon\left(  y\right)  -\beta^{2}\right]
/\left(  2\beta\right)  z\right\}  $, which leads to the one-dimensional
Schr\"{o}dinger equation $i\phi_{z}\left(  x,z\right)  +\left(  1/2\beta
\right)  \phi_{xx}\left(  x,z\right)  =0$. If the function $\Phi_{0}\left(
y\right)  $ is chosen as the magnetic (electric) field distribution of TM
(TE) plasmonic mode of the waveguide, and the parameter $\beta=\beta
_{r}+i\beta_{i}$ is the corresponding propagation constant, $I_{2}/I_{0}%
+k_{0}^{2}\varepsilon\left(  y\right)  -\beta^{2}=0$ and the Schr\"{o}dinger
equation for the amplitude $\psi$ is%
\begin{equation}
i\frac{\partial\psi\left(  s,\xi\right)  }{\partial\xi}+\frac{\partial^{2}%
\psi\left(  s,\xi\right)  }{\partial s^{2}}=0, \label{schrodinger}%
\end{equation}
where $s=x/x_{0}$ is the dimensionless transverse coordinate, $\xi=z/2\beta
x_{0}^{2}$ is the dimensionless complex propagation distance, and $x_{0}$ is
an arbitrary transverse scale.

For Eq. (\ref{schrodinger}), its finite energy Airy plasmon solution at the
input of $\psi\left(  s,0\right)  =\operatorname{Ai}\left(  s\right)
\exp\left(  as\right)  $ is%
\begin{align}
\psi\left(  s,\xi\right)  =  &  \operatorname{Ai}\left(  s-\xi^{2}%
+i2a\xi\right)  \exp\left[  i\left(  s\xi+a^{2}\xi-2\xi^{3}/3\right)  \right]
\nonumber\\
&  \times\exp\left(  as-2a\xi^{2}\right)  , \label{solution}%
\end{align}
where $a$ is a positive decay factor to truncate the amplitude at the negative
infinity. According to the integral representation of Airy function
\cite{Airy-function}, Eq. (\ref{solution}) can also be built using plane
waves
\begin{equation}
\psi\left(  s,\xi\right)  =\frac{1}{2\pi}\int_{-\infty}^{+\infty}\Phi\left(
k_{s},\xi\right)  e^{ik_{s}s}dk_{s}, \label{plane wave}%
\end{equation}
where
\begin{equation}
\Phi\left(  k_{s},\xi\right)  =e^{\frac{a^{3}}{3}}e^{-ia^{2}k_{s}}%
e^{-ak_{s}^{2}}e^{i\frac{k_{s}^{3}}{3}}e^{-ik_{s}^{2}\xi} \label{Fourier}%
\end{equation}
is the Fourier spectrum of the $k$-space, the cubic phase term $\exp\left(
ik_{s}^{3}/3\right)  $ is associated with the spectrum of the Airy wave, the
first Gaussian function $\exp\left(  -ak_{s}^{2}\right)  $ arises from the
exponential apodization of the beam, and the second Gaussian function
$\exp\left(  k_{s}^{2}\xi_{i}\right)  $ is nonzero with $\xi_{i}=-\beta
_{i}z/\left[  2\left(  \beta_{r}^{2}+\beta_{i}^{2}\right)  x_{0}^{2}\right]  $
because of the optical loss of graphene. From Eqs. (\ref{electric field}),
(\ref{plane wave})-(\ref{Fourier}), the $x$ and $z$ components of the
wavevector are $k_{x}=k_{s}/x_{0}$ and $k_{z}=\left(  \beta_{r}+\delta
\beta_{r}\right)  +i\left(  \beta_{i}+\delta\beta_{i}\right)  $, respectively,
where $\delta\beta_{r}=-k_{x}^{2}\beta_{r}/\left[  2\left(  \beta_{r}%
^{2}+\beta_{i}^{2}\right)  \right]  $, and $\delta\beta_{i}=k_{x}^{2}\beta
_{i}/\left[  2\left(  \beta_{r}^{2}+\beta_{i}^{2}\right)  \right]  $. To
insure the validity of the quasi-TM or quasi-TE mode condition, the wavevector
components must satisfy $\left\vert k_{x}\right\vert \ll\beta_{r}$,
$\left\vert \delta\beta_{r}\right\vert \ll\beta_{r}$, and $\left\vert
\delta\beta_{i}\right\vert \ll\beta_{i}$. Given the Gaussian spectrum of the
Airy plasmon in Eq. (\ref{Fourier}), the three conditions reduce to%
\begin{equation}
\sqrt{a+\frac{\beta_{i}z}{2\left(  \beta_{r}^{2}+\beta_{i}^{2}\right)
x_{0}^{2}}}\beta_{r}x_{0}\gg1. \label{condition_1}%
\end{equation}
Since Eq. (\ref{condition_1}) should be fulfilled for arbitrary propagation
distance $z$, we let%
\begin{equation}
\sqrt{a}\beta_{r}x_{0}\gg1. \label{condition_1_reduce}%
\end{equation}
Besides, the paraxial approximation holds provided that $\left\vert
\partial^{2}\psi/\partial z^{2}\right\vert \ll\left\vert 2i\beta\partial
\psi/\partial z\right\vert $, namely $\left\vert \delta\beta\right\vert
\ll2\left\vert \beta\right\vert $. This approximation also holds, if Eq.
(\ref{condition_1_reduce}) is valid.

\begin{figure}[ptb]
\centering
\vspace{0.0cm} \includegraphics[width=8cm]{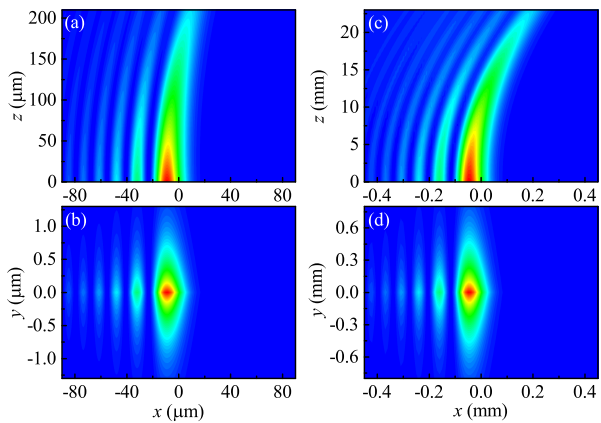} \vspace{-0.0cm}%
\caption{(Color Online) Left panel: The magnetic field intensity distribution of
the quasi-TM Airy plasmons at (a) the $xz$ plane with $y=0$ and (b) the $xy$
plane with $z=0$, respectively, where $a=0.1$, $f=10$ THz, $x_{0}=10$ $\mu$m, and
$\mu_{c}=0.8$ eV. Right panel:
The electric field intensity distribution of the quasi-TE Airy plasmons at (c)
the $xz$ plane with $y=0$ and (d) the $xy$ plane with $z=0$, respectively,
where $a=0.1$, $f=100$ THz, $x_{0}=50$ $\mu$m, and $\mu_{c}=0.1$ eV.}%
\label{field}%
\end{figure}

Considering the optical loss of graphene and according to Eq. (\ref{solution}%
), the parabolic self-deflection experienced by the Airy plasmons during
propagation can be estimated as
\begin{equation}
z^{2}=4\beta_{r}^{2}x_{0}^{3}x.
\label{parabolic}%
\end{equation}
Taking
$L_{a}=1/2\beta_{i}$ as the analytically estimated propagation length
of Airy plasmons
\cite{OL35-2082}, the transverse displacement of the beam at the propagation
length can be calculated analytically as
\begin{equation}
\Delta x_{a}(z=L_{a})=\frac{1}{16\beta_{r}^{2}\beta_{i}^{2}x_{0}^{3}}.
\label{displacement}%
\end{equation}
To realized the Airy plasmons in graphene-based waveguides, Eq.
(\ref{condition_1_reduce}) must be fulfilled to insure the validity of the
paraxial approximation. Since the decay factor $a$ is usually small to avoid
excessively changing the non-diffracting behavior of Airy beams, the real part
of propagation constant $\beta_{r}$ and the transverse scale $x_{0}$ must be
large enough. However, from Eq. (\ref{displacement}), the decrease of the
transverse displacement $\Delta x_{a}$ would be a challenge for the detection
and measurement of Airy plasmons experimentally. A contradiction exists
between the validity of the paraxial approximation and a large enough
transverse displacement.

From Eq. (\ref{solution}), the decay factor $a$ imposes an attenuation to the
propagation of Airy plasmons, and $\xi_{i}$ introduces extra exponential
terms. This imposes errors to the analytical results $L_{a}$ and $\Delta
x_{a}$. Thus we can also calculate the propagation length of Airy plasmons
numerically, and compare it with $L_{a}$. The propagation length
$L_{n}$ is defined as the distance where the input power $P=\int_{-\infty
}^{+\infty}\left\vert \Phi\left(  x,y=0,z\right)  \right\vert ^{2}dx$
decreases to $e^{-1}P_{0}$ along the propagation
direction, where $P_{0}=\int_{-\infty
}^{+\infty}\left\vert \Phi\left(  x,y=0,z=0\right)  \right\vert ^{2}dx$
is the initial input power.
Accordingly, the
transverse displacement $\Delta x_{n}$ can be defined as the displacement of the
maximum field intensity during propagation.
Since the solution of Airy plasmons is assumed to be a perturbation of
the plasmonic mode, our model is valid if the analytical propagation length
$L_{a}$ and transverse displacement $\Delta x_{a}$ are approximately equal to the numerical
propagation length $L_{n}$ and transverse displacement $\Delta x_{n}$,
respectively.


First, we focus on the quasi-TM Airy plasmons. Figs. \ref{field}(a) and (b)
show the magnetic field intensity distributions at the $xz$ plane with $y=0$
and the $xy$ plane with $z=0$, respectively, where the parameters are $a=0.1$,
$f=10$ THz, $x_{0}=10$ $\mu$m, and $\mu_{c}=0.8$ eV to insure the validity of
Eq. (\ref{condition_1_reduce}). The field intensity at the $xz$ plane exhibits
the self-deflection behavior of Airy beam, and the intensity at the $xy$ plane
is governed by the localized TM plasmonic mode of the graphene-based
waveguide. Since the TM mode is highly localized along the graphene surface,
the width of the main lobe is much larger than its height. Meanwhile, although
the surface conductivity of monolayer graphene is almost purely imaginary with
$\sigma_{g}=0.006+1.498i$ mS, the energy attenuation experienced by TM Airy
plasmons is large and the propagation length is quite limited. For the
parameters in Figs. \ref{field}(a) and (b), the propagation length is
$L_{n}=170.5$ $\mu$m ($L_{a}=175.4$ $\mu$m), and the corresponding transverse
displacement at the propagation length is $\Delta x_{n}=12.1$ $\mu$m
($\Delta x_{a}=12.9$ $\mu$m),
where the analytical result is nearly equal to the numerical result.
If we take $2x_{0}$ as a measure of the width of the main lobe, the beam is
only displaced by half width approximately.

Similarly, for the quasi-TE Airy plasmons, Figs. \ref{field}(c) and (d) show
the electric field intensity distribution at the $xz$ plane with $y=0$ and
$xy$ plane with $z=0$, respectively, where the parameters are $a=0.1$,
$f=100$ THz, $x_{0}=50$ $\mu$m, and $\mu_{c}=0.1$ eV to insure the validity of Eq.
(\ref{condition_1_reduce}). The field intensity at the $xz$ plane also
exhibits the self-deflection behavior of Airy beam, while the intensity at the
$xy$ plane is governed by the localized TE plasmonic mode of
graphene-based waveguide. Since the TE mode is weakly localized along the
graphene surface, the width of the main lobe is much smaller than its height.
Meanwhile, the energy attenuation experienced by TE Airy plasmons is small, and
the propagation length is quite large due to the weak localized field. For the
parameters in Figs. \ref{field}(c) and (d), the propagation length is
$L_{n}=28.1$ mm ($L_{a}=29.6$ mm), and the corresponding transverse displacement
at the propagation length is $\Delta x_{n}=0.4$ mm
($\Delta x_{a}=0.4$ mm), which is about 4 times of
the width of the main lobe.

By comparing the quasi-TM Airy plasmons with the quasi-TE Airy plasmons, we
can conclude that the quasi-TE Airy plasmons have a larger transverse
displacement because of the weak confinement and low propagation loss.
This is favourable to the realization of Airy beams in
experiments. However, as shown in Fig. \ref{modes}, the propagation constant
of TM plasmonic mode is sensitive to the chemical potential of graphene, which
can be tuned by the chemical doping and/or a gate voltage \cite{JAP103-064302}.
Thus the
propagation of TM Airy plasmons can be steered by the chemical potential externally.

\section{Beam steering}

\begin{figure}[ptb]
\centering
\vspace{0.0cm} \includegraphics[width=8cm]{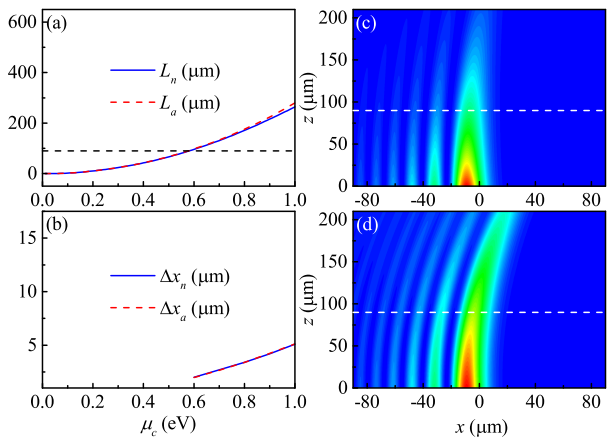} \vspace{-0.0cm}%
\caption{(Color Online) Left panel:
(a)-(b) The (a) propagation length and (b) transverse displacement at
$z=90$ $\mu$m of quasi-TM Airy plasmons in the monolayer graphene based
waveguide for different values of chemical
potential, where $L_{n}$ and $\Delta x_{n}$ are numerical results,
and $L_{a}$ and $\Delta x_{a}$ are analytical results.
Right panel: The magnetic field intensity
distributions of the quasi-TM Airy plasmons at the $xz$ plane with $y=0$
for (c) $\mu_{c}=0.6$ eV, and (d) $1.0$ eV, respectively.
The dashed black line in (a) and white lines in (c)-(d) indicate $z=90$ $\mu$m.
The parameters are $a=0.1$, $f=10$ THz, and $x_{0}=10$ $\mu$m.}%
\label{steering}%
\end{figure}

To steer the self-deflection behavior of quasi-TM Airy plasmons,
the chemical potential of monolayer graphene can be tuned externally.
As shown in Fig. \ref{steering} (a), the propagation length
of Airy plasmons changes effectively when the chemical potential
changes, where the parameters are $a=0.1$, $f=10$ THz, and
$x_{0}=10$ $\mu$m to insure the validity of Eq. (\ref{condition_1_reduce}).
Note the analytical result $L_{a}$ calculated from Eq. (\ref{parabolic})
agrees well with the numerical result $L_{n}$.
When the chemical potential is small, the imaginary part of the
propagation constant $\beta_{i}$ is large and the corresponding propagation length
is small. This is unfavorable for the propagation of Airy plasmons.

To evaluate the tunability of self-deflection behavior, we require that
the propagation length should be larger than $90$ $\mu$m, which is 3
times of the wavelength in free space.
Under this requirement, the chemical potential is within
$0.6$ eV $\le \mu_c \le 1.0$ eV,
and the transverse displacement of Airy plasmons at $z=90$ $\mu$m varies
from $2.0$ $\mu$m to $5.1$ $\mu$m, as shown in Fig. \ref{steering}(b).
Fig. \ref{steering} (c)-(d) show the magnetic field intensity distribution
of the quasi-TM Airy plasmons in the monolayer graphene based
waveguide for $\mu_{c}=0.6$ eV, and
$1.0$ eV, respectively. Clearly, the transverse displacement is changed
effectively, which is promising for the detection, sensing, and other
applications.

\begin{figure}[ptb]
\centering
\vspace{0.0cm} \includegraphics[width=8cm]{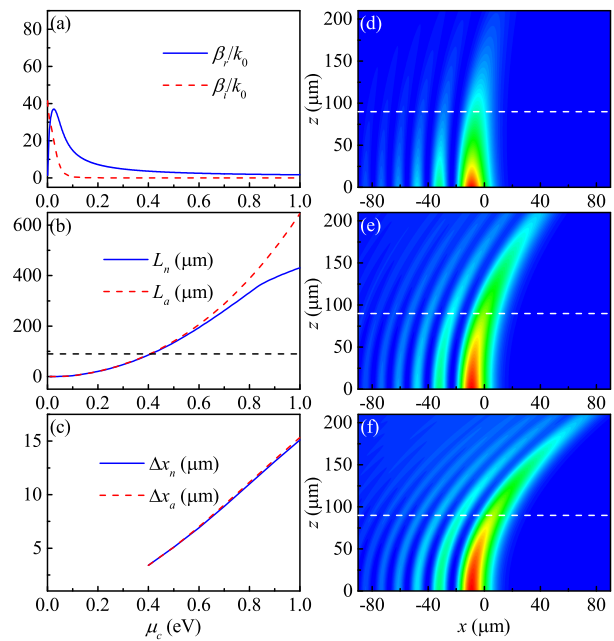} \vspace{-0.0cm}%
\caption{(Color Online) Left panel: (a) The dependence between the
propagation constant and the chemical potential for the TM plasmonic
mode in the bilayer graphene based waveguide, where $\beta_{r}$ is the
real part of propagation constant, and $\beta_{i}$ is the imaginary part.
(b)-(c) The (b) propagation length and (c) transverse displacement at
$z=90$ $\mu$m of quasi-TM Airy plasmons for different values of the chemical
potential, where $L_{n}$ and $\Delta x_{n}$ are numerical results,
and $L_{a}$ and $\Delta x_{a}$ are analytical results.
Right panel: The magnetic field intensity
distributions of the quasi-TM Airy plasmons at the $xz$ plane with $y=0$
for (d) $\mu_{c}=0.4$ eV, (e) $0.7$ eV, and (f) $1.0$ eV, respectively.
The dashed black line in (b) and white lines in (d)-(f) indicate $z=90$ $\mu$m.
The parameters are $a=0.1$, $f=10$ THz, and $x_{0}=10$ $\mu$m.}%
\label{multilayer}%
\end{figure}

In the above discussion, we always use monolayer graphene. However,
multilayer graphene is also important for the study of Airy plasmons.
First, it is hard to fabricate the strict monolayer graphene in experiments,
and it is very likely that the used graphene is not monolayer in practical
applications. Second, the plasmonic mode supported by the multilayer
graphene has a longer propagation length \cite{APL101-111609,JLT32-3597},
and it is likely to increase
the propagation length and transverse displacement of Airy plasmons by
increasing the number of layers of graphene. Considering the two factors,
in the following, we study the Airy plasmons and try to improve the
tunability of beam steering based on multilayer graphene.

The surface conductivity of multilayer graphene can be calculated as
$N\sigma_{g}$ for $N<6$ \cite{LPR8-291,NL7-2711,nnano7-330},
where $N$ is the number of layers, and $\sigma_{g}$
is the surface conductivity of monolayer graphene. For the calculation
of plasmonic mode and Airy plasmons based on multilayer graphene,
we only need to replace $\sigma_{g}$ for the monolayer graphene
with $N\sigma_{g}$. We consider the bilayer graphene with $N=2$
as an example.
As shown in Fig. \ref{multilayer} (a), both the
real and imaginary parts of propagation constant of TM plasmonic
mode in the bilayer graphene based waveguide decrease compared with the results
shown in Fig. \ref{modes}(b). Thus, for the same parameters of
$a=0.1$, $f=10$ THz, and $x_{0}=10$ $\mu$m,
Airy plasmons based on bilayer graphene have
a larger propagation length, as shown in Fig. \ref{multilayer} (b).
Meanwhile, the transverse displacement at a same propagation distance
would also increase according
to Eq. (\ref{parabolic}). Note in Fig. \ref{multilayer} (b),
the analytical results deviate from the numerical results for large
values of chemical potential. This is due to the decrease of
$\sqrt{a}\beta_{r}x_{0}$ in the paraxial approximation condition
of Eq. (\ref{condition_1_reduce}).

Similarly, to evaluate the tunability of self-deflection behavior, we also
require that the propagation length should be larger than
$90$ $\mu$m. Under this requirement, the chemical potential of
bilayer graphene is within $0.4$ eV $\le \mu_c \le 1.0$ eV,
and the transverse displacement of Airy plasmons at $z=90$ $\mu$m varies
from $3.4$ $\mu$m to $15.2$ $\mu$m.
The range of transverse displacement is almost 4 times of that for the Airy
plasmons based on monolayer graphene.
Fig. \ref{multilayer} (d)-(f) show the magnetic field intensity distribution
of the quasi-TM Airy plasmons in the bilayer graphene based waveguide
for $\mu_{c}=0.4$ eV, $0.7$ eV, and $1.0$ eV, respectively.
Compared with the Airy plasmons
based on monolayer graphene, Airy plasmons based on bilayer graphene shows a
larger tunability, where the self-deflection behavior can be tuned more
effectively. Thus the multilayer graphene provides a better platform to
the beam steering compared with the monolayer graphene.

\section{Conclusion}

In conclusion, we derive an analytical model under the paraxial approximation
to study the Airy plasmons in graphene-based waveguides.
Both the quasi-transverse-magnetic (TM) and
quasi-transverse-electric (TE) Airy plasmons can be supported,
where the quasi-TE Airy plasmons have larger propagation length and
transverse displacement, and the quasi-TM Airy plasmons show a
better tunability to steer the self-deflection behavior.
Moreover, for the quasi-TM Airy plasmons, the propagation length
and the range of transverse displacement can be increased,
if multilayer graphene is used in the graphene-based waveguides.
Besides the metals, graphene provides an
additional platform to investigate the propagation of Airy plasmons
and to design various plasmonic devices.

\section{Acknowledgment}

This work was sponsored by the National Natural Science Foundation of China
under Grants No. 61322501, No. 61574127, and No. 61275183, the Top-Notch Young
Talents Program of China, the Program for New Century Excellent Talents
(NCET-12-0489) in University, the Fundamental Research Funds for the Central
Universities, and the Innovation Joint Research Center for
Cyber-Physical-Society System.

\end{document}